\def\vep{\varepsilon}

\def\al{\alpha}

\def\be{\begin{equation}}
\def\ee{\end{equation}}
\def\bea{\begin{eqnarray}}
\def\eea{\end{eqnarray}}

\def\nostrocostrutto#1\over#2{\mathrel{\mathop{\kern 0pt \rlap 
  {\raise.2ex\hbox{$#1$}}}
  \lower.9ex\hbox{\kern-.190em $#2$}}}
\def\lsim{\nostrocostrutto < \over \sim}   

\newcommand{\ba}{\begin{eqnarray}}
\newcommand{\ea}{\end{eqnarray}}

\documentstyle[12pt,epsfig]{article}

\def\@sect#1#2#3#4#5#6[#7]#8{\ifnum #2>\c@secnumdepth
     \let\@svsec\@empty\else  
     \refstepcounter{#1}\edef\@svsec{\csname the#1number\endcsname}\fi
     \@tempskipa #5\relax
      \ifdim \@tempskipa>\z@
        \begingroup #6\relax
          \@hangfrom{\hskip #3\relax\@svsec}{\interlinepenalty \@M #8\par}%
        \endgroup
       \csname #1mark\endcsname{#7}\addcontentsline
         {toc}{#1}{\ifnum #2>\c@secnumdepth \else
                      \protect\numberline{\csname the#1\endcsname}\fi
                    #7}\else
        \def\@svsechd{#6\hskip #3\relax  
                   \@svsec #8\csname #1mark\endcsname
                      {#7}\addcontentsline
                           {toc}{#1}{\ifnum #2>\c@secnumdepth \else
                             \protect\numberline{\csname the#1\endcsname}\fi
                       #7}}\fi
     \@xsect{#5}}

\advance\textheight by \topskip
 \advance\footskip by 6mm
\def\vep{\varepsilon}

\def\al{\alpha}

\renewenvironment{abstract}{\if@twocolumn \section*{ABSTRACT}\else
\begin{quote}\begin{center}%
\bf \abstractname\par
\end{center}\vskip 0.5ex\fi}
{\if@twocolumn\else\end{quote}\vskip 3ex\fi}

\catcode`@=11
\newcount\@tempcntc
\def\@citex[#1]#2{\if@filesw\immediate\write\@auxout{\string\citation{#2}}\fi
  \@tempcnta\z@\@tempcntb\m@ne\def\@citea{}\@cite{\@for\@citeb:=#2\do
    {\@ifundefined
       {b@\@citeb}{\@citeo\@tempcntb\m@ne\@citea\def\@citea{,}{\bf ?}\@warning
       {Citation `\@citeb' on page \thepage \space undefined}}%
    {\setbox\z@\hbox{\global\@tempcntc0\csname b@\@citeb\endcsname\relax}%
     \ifnum\@tempcntc=\z@ \@citeo\@tempcntb\m@ne
       \@citea\def\@citea{,}\hbox{\csname b@\@citeb\endcsname}%
     \else
      \advance\@tempcntb\@ne
      \ifnum\@tempcntb=\@tempcntc
      \else\advance\@tempcntb\m@ne\@citeo
      \@tempcnta\@tempcntc\@tempcntb\@tempcntc\fi\fi}}\@citeo}{#1}}
\def\@citeo{\ifnum\@tempcnta>\@tempcntb\else\@citea\def\@citea{,}%
  \ifnum\@tempcnta=\@tempcntb\the\@tempcnta\else
   {\advance\@tempcnta\@ne\ifnum\@tempcnta=\@tempcntb \else \def\@citea{--}\fi
    \advance\@tempcnta\m@ne\the\@tempcnta\@citea\the\@tempcntb}\fi\fi}
\catcode`@=12
\begin{document}
\noindent
\rightline{MPI-PhT/96-114}
\rightline{November, 1996}
\vfill
\begin{center}
{\Large\bf 
       Scaling and scale breaking phenomena in QCD jets}
\end{center}
\vfill
\begin{center}
WOLFGANG OCHS 
\end{center}

\begin{center}
\mbox{ }\\
     \ {\it Max-Planck-Institut f\"ur Physik \\
(Werner-Heisenberg-Institut)\\
F\"ohringer Ring 6, D-80805 Munich, Germany} 
\end{center}

\vfill
\begin{abstract}
The perturbative QCD approach to multiparticle production assuming Local
Parton Hadron Duality (LPHD) and some recent results
are discussed. Finite asymptotic scaling limits
are obtained for various observables, after an appropriate rescaling, in the
Double Logarithmic Approximation (DLA).
Non-asymptotic corrections are also known in some cases. The DLA applies also
to very soft particle production where energy conservation constraints can
be neglected.
In this region the particle density follows rather
well a scaling behaviour over the full energy range explored so far in
$e^+e^-$ annihilation.
\end{abstract}
\vfill
Lecture at the XXXVI Cracow School of Theoretical Physics, Zakopane, Poland,
June 1996.
\newpage

\section{Introduction}
The study of the intrinsic structure of particle jets produced
in hard collisions continues to be an active field of research.
The interest is directed in elaborating and testing specific
predictions of perturbative QCD on the parton cascade evolution
and secondly, to investigate the hadronization process which cannot
be treated within a perturbative scheme. A reduction of the flexiblity
of the models involved and a deeper understanding of the phenomenological
aspects of the confinement process is an important aim of this research.

The most popular models for particle production in 
hard collision processes  are based on a
primary hard partonic sub-process which is accompagnied by gluon bremsstrahlung.
The evolution of the partonic jets is derived in perturbation theory and is
terminated at a scale of around 1~GeV; thereafter, non-perturbative
processes take over and the final hadronic particles, often through
intermediate resonances, are produced, for example, by a string
mechanism
\cite{JETSET}
or through  cluster formation
\cite{HERWIG}.

Another approach is based on the concept of ``Local Parton Hadron
Duality'' (LPHD)
\cite{adkt}.
It has been observed at first that the hadronic energy spectra are
rather well represented by the parton spectra themselves -- without an
additional hadronization phase -- provided the cut-off of the parton cascade
is lowered to a  value around 250~MeV, of the order of the hadronic
masses. This general idea has been applied to various other observables; 
the theoretical calculations are based in the simplest case on the
Double Log Approximation (DLA)
\cite{dfk1,bcm},
which provides the high energy limit, or the Modified Leading
Log Approximation (MLLA)
\cite{MLLA}
which includes finite energy corrections which are usually essential
to obtain quantitative agreement with experiment at present
energies
\cite{dkmt}.
Recent experimental results from LEP, HERA and TEVATRON gave further
support to this approach
\cite{ko}.
Although a justification of the model is not yet available
at a fundamental level the related phenomenology is quite attractive
because of its intrinsic simplicity with very few parameters.
Also the analytical computations allow the derivation of scaling laws and
the systematics of their violation which provides an important insight into
the structure of the theory.
On the other hand, it is clear that this model cannot compete with
the standard hadronization models in the description of 
the various details of the
final state like the production of different 
particle species or resonances. It has so far been applied successfully
for suitably averaged quantities.

In this presentation we summarize, how scaling and scale breaking
predictions obtained from analytical calculations compare with experiment.

\section{Basic ingredients of analytical calculations}
We consider high energy collisions which involve  a hard subprocess.
The colour charges of the primarely  
produced partons are the sources of subsequent gluon
bremsstrahlung which leads to the partonic jets.
The subprocess is described by the corresponding matrix element. 
The  gluon bremsstrahlung at small angles $\delta$ with energy $E$ 
 off the primary hard parton of type $A$ ($A = q, g$)
with momentum $P$ is given by 

\begin{equation}
  dn_A = \frac{C_A}{N_C} \gamma_0^2(k_T) \frac{d \delta}{\delta} \frac{dE}{E},
 \qquad
  \gamma_0^2(k_T) = \frac{2N_C \alpha_s(k_T)}{\pi}
                  = \frac{\beta^2}{\ln(k_T/\Lambda)}, 
  \qquad k_T \ge Q_0
\label{gluon}
\end{equation}
where $k_T \approx E \delta$, $\beta^2 = 4 N_C/b$,
$b \equiv (11 N_C - 2n_{f})$/3 with $N_C$, $n_f$
the numbers of colours and flavours, also $C_g$ = $N_C$, $C_q$ = 4/3.
Inside the cascade the soft gluons are coherently produced from all
harder partons. For azimuthally averaged quantities the consequences
of the coherence effect can be taken into account by the angular 
ordering prescription \cite{AO} which requires
 the angles of subsequent gluon emissions to be in decreasing order.

The multiparticle properties of the jet can be discussed
conveniently by using the generating functional \cite{ahm}
$Z_A(P, \Theta;{u(k)})$. Here $P$ and $\Theta$ denote the initial parton
momentum and 
opening angle of the jet, and $u(k)$ is a profile function for
particle momentum $k$. The functional is constructed from all
the exclusive final states. Then the inclusive densities
can be obtained by functional differentiation with respect to 
 the profile function
$u(k)$
\begin{equation}
 \rho^{(n)} (k_1,...,k_n)= \delta^n Z\{u\}/\delta u(k_1)...
\delta u(k_n)\mid_{u=1}.      \label{dn}
\end{equation}
Properties of these densities can be obtained from the
evolution equation for $Z$ which relates the functional
at scales $P, \Theta$ to the one at lower scales according to the
``decay'' $A \to BC$. In MLLA accuracy this evolution equation
is given by \cite{MLLA}
\begin{eqnarray}
\frac{d}{d \: \ln \: \Theta} \: Z_A (P, \Theta) & = &
\frac{1}{2}
\; \sum_{B,C} \; \int_0^1 \; dz \nonumber \\
& & \label{evz}\\
 \; &\times &\frac{\alpha_s (k_\perp^2)}{2 \pi} \: \Phi_A^{BC} (z)
\left [Z_B (zP, \Theta) \: Z_C ((1 - z)P, \Theta) \: - \: Z_A
(P,\Theta) \right ] \nonumber
\end{eqnarray}
where $\Phi_A^{BC}(z)$ denotes the DGLAP splitting functions.
The initial condition of the evolution is given by
\begin{equation}
Z_A (P, \Theta; \{ u \})|_{P \Theta \: = \: Q_0} \; = \; u_A
(k  
\: = \: P),
\label{ic}
\end{equation}
i.e. at threshold there is only the primary parton.

These equations take into account energy conservation 
by choosing the proper arguments of $Z$. One can obtain
for various observables $O$ 
analytic solutions which correspond to the summation
of the perturbative series in
leading double logarithmic order, i.e. the summation of the terms 
$\alpha_s^n L^{2n}$ with a large logarithm  $L$.  Also results with resummed
next-to-leading order terms $\alpha_s^n L^{2n-1}$ are available in some
cases. 
At high energies $O \sim \exp \int^Y \gamma(\alpha_s(y))\; dy$
where the anomalous dimension $\gamma$ has the expansion 
$\gamma \sim  \sqrt{\al_s} +\alpha_s +\ldots $ 
The leading term is refered to as the DLA, 
the next-to-leading one as the MLLA result. The evolution equation (\ref{evz})
yields the complete results for the first two terms in this 
$\sqrt{\al_s}$ expansion.
These leading terms are not 
sufficient, however, to satisfy the initial condition (\ref{ic}),
this is only possible by using the full result from the summed 
 perturbative series. 
From (\ref{evz}) one can obtain the evolution equations for
particle densities by appropriate differentiation (\ref{dn}). Therefore
this equation is the basic tool for deriving the multiparticle
properties of a jet analytically.

For very high energies the small $z$ contributions dominate (fixed scales
$Q_0$ and $\Lambda$), then    
one can neglect the recoil effects and 
approximate 
$1 - z \approx 1$ in the argument of $Z$ in (\ref{evz}). Furthermore, in the
high energy limit it is sufficient to include the most singular terms
of the splitting functions $\Phi_A^{Ag}\sim 1/z$, in particular, one can
neglect the production of quark pairs in the cascade with nonsingular
splitting function. In this case Eq. (\ref{evz}) simplifies and can be
integrated using the initial condition (\ref{ic}) to 
\ba
Z_A(P,\Theta,{u}) = u(P) \exp\left(\int_\Gamma
dn_A[u(E)Z_g(E,\delta)-1]\right)
\label{dlaz}
\ea
with integration measure from (\ref{gluon}) and boundary $\Gamma$ which
takes into account 
the angular ordering constraint
$\delta<\Theta$ and the $k_T$ cutoff $E\delta>Q_0$. This is the evolution
equation in DLA accuracy appropriate for the high energy asymptotics.

For the energy spectra \cite{dfk2} and a large class of angular
correlations \cite{ow} one can derive from (\ref{dlaz}) by functional
differentiation and appropriate partial integration an evolution equation
of the type 
\be
h_n(\delta,\Theta,P)= d_n(\delta,P) +\int \frac{dK}{K}\int
\frac{d\psi}{\psi}\; \gamma_0^2\; h_n(\delta, \psi,K)
\label{hnevol}
\ee
where $h_n$ denotes generically one such distribution or correlation of order
$n$ and $d_n$ the appropriate initial condition. The singularities in the
kernel are regularized by the $k_T$ cut-off. A nonsingular evolution
equation is obtained by changing to logarithmic momentum 
and angular variables. 

\section{High energy asymptotics}
The high energy behaviour can be obtained from (\ref{dlaz}). As will be
shown in several cases, 
the observable quantities, after appropriate rescaling, approach a 
finite scaling limit.

\subsection{Multiplicity distribution}
A well known example of such behaviour is the ``KNO-scaling''
\cite{KNO,poly} of the multiplicity distribution

\begin{equation}
 <n> P_n(s) = f(\psi), \qquad \psi = n/<n(s)>.
\label{kno}
\end{equation}
Here $f(\psi)$ is the high energy limit of  the probability
$P_n$
to produce $n$ particles at $cms$ energy $\sqrt{s}$,
rescaled by the average multiplicity $<n>$.
This scaling law has been derived in the DLA for the partons 
of QCD \cite{bcm1}
but holds more generally for a large class of branching processes
\cite{poly,or}. Specifically for QCD one obtains also an explicit prediction
for the
function $f(\psi)$ or the normalized factorial moments
$F^{(k)} = <n(n-1) ... (n - k + 1)>$/$<n>^k$ of the multiplicity distribution.
For example, for $e^+e^-$ annihilation one finds
$F^{(2)} = \frac{11}{8}$.
This prediction is infrared safe, i.e.~independent of the
cut-off $Q_0$. Furthermore QCD predicts the approach to the asymptotic
limit, so for the same quantity one obtains
with inclusion of the next-to-leading correction (MLLA)
\cite{mw}

\begin{equation}
 F^{(2)} = \frac{11}{8}(1-\frac{4255}{1782 \sqrt{6\pi}} \sqrt\al_s)
\label{f2mom}
\end{equation}
(for $n_f$ = 5) which turns out to be large (about 30\%); if yet higher order
corrections are included the result fits the experimental data
\cite{vietri}. The energy dependence of $F^{(2)}$ in (\ref{f2mom}) 
is very weak ($\sim 1/\sqrt{\ln s}$) thus simulating the  scaling behaviour 
observed at present energies. The ultimate KNO scaling function according to 
 (\ref{kno}) or (\ref{f2mom})
is broader and will be approached only at much higher energies
than available today and in the near future. 

\subsection{Momentum spectra}
Spectra in the rescaled Bjorken or Feynman momentum variables
$x = p/P$ do not scale in QCD. Rather the distributions of 
certain rescaled
logarithmic variables approach a finite asymptotic limit in the
DLA. Such  scaling properties have been discussed recently in some
detail for angular correlations
\cite{ow}
(see below). For the energy  spectra a scaling limit 
of this type has been suggested
already some time ago
\cite{dfk2} and one obtains in logarithmic variables after rescaling

\begin{equation}
  \frac{\ln {dn}/{d\xi}}{\ln<n>} = f(\zeta), 
 \qquad \zeta = \frac{\xi}{Y+\lambda}
\label{zeta}
\end{equation}
where $\xi = \ln(P/E) = \ln(1/x)$ for a particle of energy $E$ in a jet with
primary parton momentum $P$, $Y = \ln(P\Theta/Q_0)$ and
$\lambda = \ln(Q_0/\Lambda)$.
The function $f(\zeta)$ has an approximately Gaussian shape, the
so-called ``hump-backed plateau''
\cite{dfk1,bcmm}. Again, the approach to this limit is rather slow, 
for example, the maximum of the spectrum occurs at
(for $Q_0 \approx \Lambda$)
\begin{equation}
 \xi^* = Y(\frac{1}{2} + \sqrt{\frac{c}{Y}} -\frac{c}{Y}) + 0.1
\label{ximax}
\end{equation}
where $c = 0.2915$ for $n_f = 3$. The leading DLA term gives the
asymptotic limit $\zeta^* \to \frac{1}{2}$, the next two
terms the high energy corrections
\cite{dktint}, 
the last term a numerical estimate of the remaining contributions
applicable in the present energy range
\cite{klo1}. According to the LPHD hypothesis this prediction at the parton
level can be compared directly to the hadronic observable.
The rescaled quantity $\zeta^*=\xi^*/Y$ for charged hadrons 
is shown in Fig.~1 up to LEP-1.5 energies in comparison
with the prediction
(\ref{ximax}). The cut-off parameter  $Q_0=0.270$ GeV
is taken from a global fit to the
moments of the distribution from $cms$ energies 3 to 91 GeV in $e^+e^-$
annihilation \cite{lo}.
The  data in Fig.~1 closely follow the MLLA prediction (\ref{ximax}) 
which very slowly approaches the asymptotic DLA limit
$\zeta^*=\frac{1}{2}$.

\subsection{Angular correlations}
There was a lot of interest in the last years in the study and
interpretation of angular correlations
\cite{dkw},
which was triggered by the suggestion
\cite{bp},
such correlations could be power behaved
at high resolution (``intermittency'').
Such power behaviour is expected, for example, for
selfsimilar branching processes, so it applies to
QCD to the extent that the running of the coupling is
neglected
\cite{ow1}.

The observables which are considered in the analytical
QCD calculations \cite{ow,dd,bmp} 
are the two particle correlation density between
two particles $\rho^{(2)} (\vartheta_{12}, \Theta)$ in the forward cone
of half angle $\Theta$, the factorial or cumulant multiplicity
moments $F^{(n)}$ and $C^{(n)}$ for particles between two cones
at angles $\Theta - \delta$ and $\Theta + \delta$
around the jet axis or in a cone of angular size $\delta$ in
direction $\Theta$ with respect to the jet axis. According
to the volume of phase space $\delta^D$ these two configurations
are refered to by their dimensions $D = 1$ und $D = 2$. One finds the
results on these correlations by first deriving the integral equation of the
$n$-particle correlation function from (\ref{dlaz}) and (\ref{dn}),
 and then integrates over the 
remaining variables; the resulting evolution equation 
is of the type (\ref{hnevol}) which
can be solved approximately for running $\alpha_s$ (for fixed $\alpha_s$ one
can  get often exact results).

The correlation functions of order $n$ are conventionally normalized by a
power of the multiplicity $<n> \;\sim \exp(2\beta \sqrt{\ln(P\Theta/\Lambda)})$.
These normalized correlations, after removal of certain known kinematic
phase space factors follow, after rescaling, the asymptotic angular
scaling law \cite{ow}

\begin{equation}
\frac{\ln H^{(n)}(\delta,\Theta,P)}{2n\beta\sqrt{\ln P\Theta/\Lambda}}
  \to (1-\omega(\vep, n)/n), 
  \qquad \vep = \frac{\ln \frac{\Theta}{\delta}}
   {\ln \frac{P \Theta}{\Lambda}}
\label{escal}
\end{equation}
in the rescaled logarithmic variable $\vep$ with
$0 \lsim \vep \lsim 1$. 
So the rescaled observables do not depend on the variables $\delta,\Theta$
and $P$ separately but only through the variable $\vep$.
The normalization of the $l.h.s$ of (\ref{escal}) corresponds to
$(\ln <n>)^n$. The function $\omega$ is known analytically, for large $n$ one
finds 
\be
\omega (\vep, n) =  n\sqrt{1-\vep} 
        (1-\frac{1}{2n^2}\ln (1-\vep)+\ldots )
\label{omax}
\ee
which turns out to be a good approximation already for $n=2$.

An interesting feature of these results is their universality, i.e.~the same
limit is obtained for quite different observables: 
 the correlations
$\hat{r}(\vartheta_{12}) = \rho^{(2)}(\vartheta_{12}, \Theta)/<n(\Theta)>^2$
and the normalized
moments of any order in one or two dimensions. These correlation 
functions refer actually to
particles in quite different regions of phase space.

As an example we show in Fig.~2 the rescaled normalized two particle
density $\hat{r}(\vartheta_{12}, \Theta)$ 
as obtained from the DELPHI collaboration
\cite{bm} which is rather well approximated by  
the asymptotic prediction from the DLA.
The data are in good agreement with the Monte Carlo calculations at the same
energy at either parton or hadron level. Monte Carlo results
at a  much higher energy show a
similar behaviour in agreement with the scaling prediction
(\ref{escal}). 
An observable which projects out the 
genuine 2-particle correlations more effectively from the 
uncorrelated background is the ``correlation integral" \cite{corint,ow}
$r(\vartheta_{12})=
\rho^{(2)}(\vartheta_{12})/\rho^{(2)}_{norm}(\vartheta_{12})$
where the normalization corresponds to the density of relative angles 
$\vartheta_{12}$ of particles from different jets. This quantity has been
measured as well \cite{bm} and the predicted angular scaling law
(\ref{escal}) for $r(\vartheta_{12})$
has been verified for different jet opening angles
$\Theta$.

An uncertainty in these comparisons, which is hard to quantify, comes from the
choice of the jet axis (taken usually as the sphericity axis). An
improvement of both theoretical and experimental results could be obtained
by using the Energy-Multiplicity-Multiplicity (EMM) definition as applied
already to the 2-particle azimuthal angle correlations \cite{dmo}.

The predictions for moments have been derived for cumulants
\cite{ow} or for factorial moments \cite{dd,bmp}
which approach the same limit asymptotically. It turns out that
the factorial moments at present energies are much closer to the
asymptotic predictions; such moments are shown in Fig.~3 
after appropriate rescaling by kinematic factors again as function of the
$\vep$-variable. 
The data show the same trend as the asymptotic DLA predictions
of (\ref{escal}). The approach of the cumulant
moments to the asymptotic results is much slower. 

Note that the
curve in Fig.~3 for $n=2$ is the same as the one in Fig.~2
according to the universality property of (\ref{escal}). 
Also the differences between
the moments of orders $n = 2$ and $n = 3$ are largely removed
after rescaling; they are  of relative 
order $1/n^2$ according to (\ref{escal}) and (\ref{omax}).

The various angular regions have quite different characteristics. 
For small $\vep$ (large relative  angles) the function
$\omega(\vep, n) \approx n -\frac{1}{2} \frac{n^2-1}{n} \vep$, which
yields a power behaviour of the moments $M^{(n)}$ (either
$C^{(n)}$ or $F^{(n)}$)

\begin{equation}
 M^{(n)}(\vartheta,\delta) \sim \left(\frac{\vartheta}{\delta}\right)
 ^{\phi_n}, \qquad \phi_n = D(n-1) - (n-\frac{1}{n})\gamma_0(P\vartheta).
\label{intsl}
\end{equation}
This is the asymptotic power law (``intermittency'') for the QCD
cascade. It applies for large relative angles where the
cascade is fully developed and reflects the selfsimilarity of the
branching process; the intermittency exponent $\phi_n$ depends on the
scale through the running $\al_s$. In case of fixed $\al_s$ the same
result is obtained with universal $\gamma_0$ parameter. 
Moving to larger $\vep$ the observables show the angular scaling law
(\ref{escal}) with non-linear function $\omega(\vep,n)$.
In this region
the results are infrared safe (i.e. do not depend on $Q_0$), they 
apparently are also not much dependent on hadronization effects
(see Fig.~2).

Moving to yet larger $\vep$ one comes to a critical angle $\vep_{crit}$ which 
separates two kinematic regimes of quite different characteristics
\cite{ow1,dmo,ow3}.
The
correlation functions have a discontinuous second derivative at this angle.
In the new  region $\vep > \vep_{crit}$ (small relative angles) the correlation
functions do depend on the cut-off $Q_0$ and they become
independent of the order $n$, contrary to the behaviour for $\vep <
\vep_{crit}$; they are given in terms of the one-particle inclusive spectrum. 
As a consequence, one expects
in this region a dependence of the particle type
($\pi\pi, KK, pp$ correlations) if the particle mass is
related to the cut-off $Q_0$. 

For fixed $\al_s$ this angle is 
given by $\vep_{crit}=n^2/(n^2+1)$ at order $n$
\cite{dmo,ow3}.
For running $\al_s$ a new scale $\Lambda$ appears and one considers
\cite{ow3} the double scaling limit of the function $\omega_n(\vep,\rho)$
with $\rho=\sqrt{\lambda/(Y+\lambda)}$ for asymptotic $Y$ at fixed
$\vep$ and $\rho$. Then again a critical behaviour at a certain angle 
 $\vep_{crit}$ is found. This limit requires also an increasing $Q_0$
and therefore does not correspond to the usual fixed 
$k_T$ cut-off. Therefore,
for finite, physical $Q_0$ the 
separation of the  two regions 
is not expected to be complete. 
It would be interesting to verify the characteristics of these two regimes
experimentally.

\section{Scaling law for soft particles}
The prediction of the hump-backed shape of the inclusive
energy spectrum in the $\xi = \ln 1/x$ variable and
its subsequent observation was an important success of
QCD in its application to multiparticle physics. The
coherence of the soft gluon emission from all harder partons
forbids the multiplication of the soft particles and one
expects nearly an energy independence of the soft particle
rate \cite{adkt}. Such a property has been pointed out to be present
indeed in the data
\cite{vakcar,lo}.

This problem has been studied recently in more detail
\cite{klo1}.
The analytical calculations both in DLA and MLLA   converge towards
the same limits independent of the $cms$ energy 
for small particle energies. In this limit the
energy conservation effects and large $z$ corrections from the splitting
functions which make up the differences between the approximations 
(\ref{evz}) and (\ref{dlaz}) can be neglected. 
If LPHD is valid towards
these low energies one expects also a scaling behaviour for the
invariant density $I_0$
of hadrons in the soft limit where the particle momentum
$p$ or rapidity $y$ and transverse momentum $k_T$ become small:
\be
I_0 = \lim_{y \to 0, p_T \to 0} E \frac{dn}{d^3p} \quad = \quad
\frac{1}{2} \lim_{p \to 0} E \frac{dn}{d^3p}.
\label{izero}
\ee
The factor $\frac{1}{2}$ in this definition takes into account that both
hemispheres are included in the limit $p \to 0$. This scaling
behaviour is a direct consequence of the coherence of the gluon
emission: The emission rate for the gluon of large wavelength
does not depend on the details of the jet evolution at smaller distances;
it is essentially determined by the colour charge of the hard initial
partons and is energy independent. The energy independent contribution comes
from the single gluon bremsstrahlung of order $\alpha_s$,
the higher order contributions 
generate the energy dependence but do not contribute in the
soft limit.

In Fig.~4 we show the experimental results on the invariant density
of charged particles
for $cms$ energies from 3 to 130 GeV in $e^+e^-$ annihilation.
An approximate energy independence of the soft limit (within about 20\%) is
indeed observed; the same is true for identified particles $\pi$,
$K$ and $p$
\cite{klo2}.
The curves in Fig.~4 represent the MLLA results, where also a
particular prescription is employed to relate the different parton and hadron
kinematics near the boundary $E\approx Q_0$
(for more details, see \cite{klo2,klo3}).
The theoretical curves show the approach to the scaling limit and 
describe well the different slopes at larger particle energies. An important
role here is played by the running $\al_s$ which provides the
strong rise towards small energies for $E<1$ GeV, for fixed $\alpha_s$ this
rise would be much weaker 
\cite{lo,klo2}.

A crucial test of the QCD-LPHD interpretation of this scaling
result is the verification of the dependence of  the
limiting densitiy $I_0$ on the
primary colour charge. This can be obtained from
$e^+e^- \to $3 jets, deep inelastic scattering or semihard
hadronic processes with gluon exchange
\cite{klo2}.

\section{Summary}
The perturbative approach to multiparticle production in
connection with the LPHD assumption represents a very
economic description of the phenomena which involves
only the parameters $Q_0$ and $\Lambda$ apart from the normalization.

The analytical treatment singles out 
the logarithmic momentum and angular variables 
which are appropriate to the description of bremsstrahlung
processes and absorb the collinear and soft divergent behaviour.
Therefore the finite asymptotic limits of various observables in the 
rescaled logarithmic variables are a direct consequence of the parton
branching process  generated by bremsstrahlung type emissions.
These scaling laws are then more specific to QCD than the KNO multiplicity
scaling which holds for a wide class of branching processes, not necessarily
of bremsstrahlung type.
These results are obtained in the DLA where energy conservation is neglected.
A noteworthy feature of angular correlations not met in energy spectra is
the occurence of a critical angle which separates two scaling regimes
with quite different characteristics.

Another scaling prediction from DLA is obtained in the soft particle
limit at finite energies where energy conservation effects can be
neglected as well. It is remarkable that perturbative QCD 
predictions work even in
such an extreme limit and this requires further investigations
with different partonic antenna patterns for confirmation.

\section*{Acknowledgements}
I would like  to thank A. Bia\l as for his inspiration of the studies
of scaling laws
in multiparticle physics and V. A. Khoze, S. Lupia and J. Wosiek for the
collaboration on the subjects of this lecture.



\begin{figure}[p]
          \begin{center}
\mbox{\epsfig{file=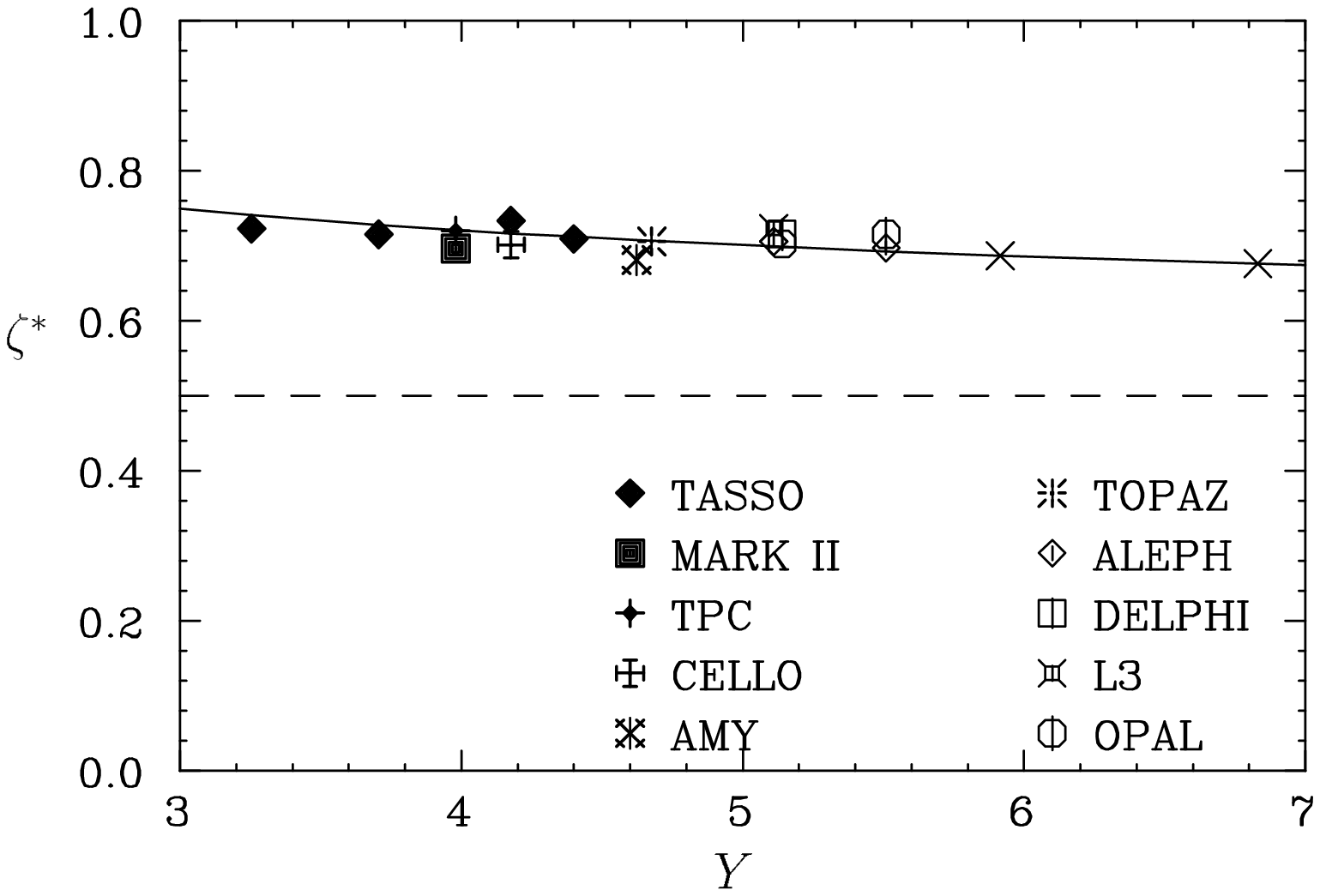,bbllx=2.8cm,bblly=14.cm,%
bburx=20.cm,bbury=26.cm,height=12.cm}}
          \end{center}
\caption{Maximum of the rescaled inclusive momentum distribution 
$\zeta^*=\xi^*/Y$  
as a function of $Y = \ln \frac{\protect\sqrt{s}}{2\Lambda}$; 
comparison between experimental
data from $e^+e^-$ annihilation  and theoretical prediction in MLLA 
numerically extracted from the shape of the Limiting Spectrum (solid line) 
for the cut-off parameter $Q_0=\Lambda$ = 270 MeV. 
Crosses mark the predictions at the $cms$ energies 
200 GeV and 500 GeV. Asymptotically, the leading DLA result $\zeta^* =
\frac{1}{2}$ is approached (see \protect\cite{klo1,lo2}).}
\end{figure}


\begin{figure}[p]
          \begin{center}
\mbox{\epsfig{file=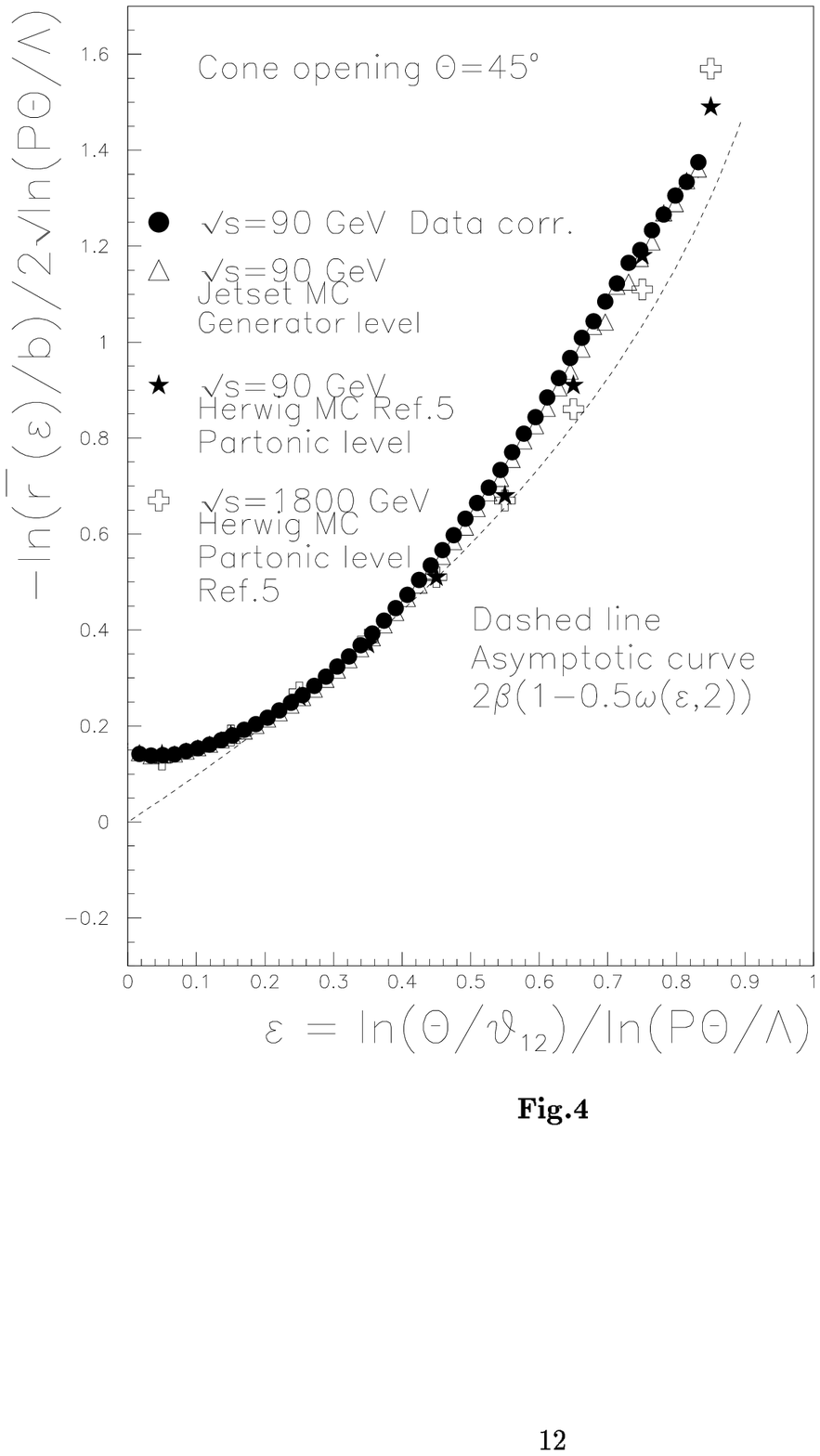,bbllx=2cm,bblly=7.cm,bburx=15cm,%
bbury=23.cm,height=16cm,clip=}}
       \end{center}
\caption{
The rescaled 2-particle angular correlation $\hat r =
\rho^{(2)}(\vartheta_{12})/<n>$ in the forward cone with half-opening
$\Theta$ as function of the
scaling variable $\epsilon$ as measured by DELPHI \protect\cite{bm}.
Also shown are the results from the JETSET and 
 HERWIG Monte Carlo's  at the parton and hadron levels at different energies.
The data show the predicted scaling behaviour and the approach to the
asymptotic DLA prediction (for $\Lambda=0.15$ GeV, with 
$b=2\beta\protect\sqrt{\ln(P\Theta/\Lambda)}$).}
\end{figure}

\newpage

%

\begin{figure}[p]
          \begin{center}
\mbox{\epsfig{file=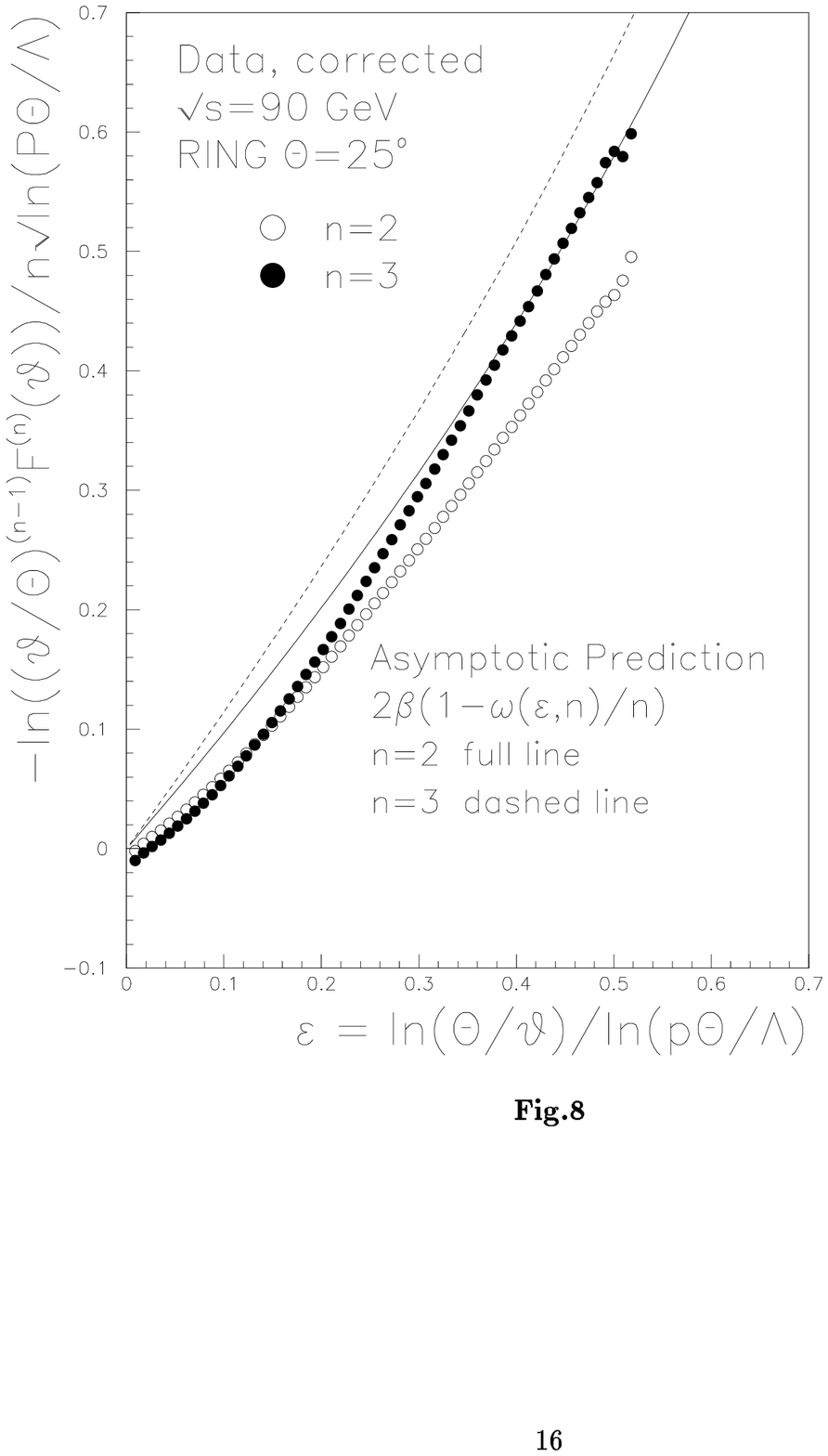,bbllx=2cm,bblly=7.cm,bburx=15cm,%
bbury=23.cm,height=16cm,clip=}}
\end{center}
\caption{Rescaled factorial multiplicity moments for particles in the ring
around the jet axis
with polar angles between $\Theta-\vartheta$ and $\Theta+\vartheta$ 
as measured by
 DELPHI \protect\cite{bm} (momentum $P=45$ GeV, $\Lambda=0.15$ GeV)
in comparison with the asymptotic DLA prediction. Note that the curve for
$n=2$ is the same as in Fig. 2 for the rescaled correlation $\hat r$.}
\end{figure}

\newpage 

\begin{figure}[p]
          \begin{center}
\mbox{\epsfig{file=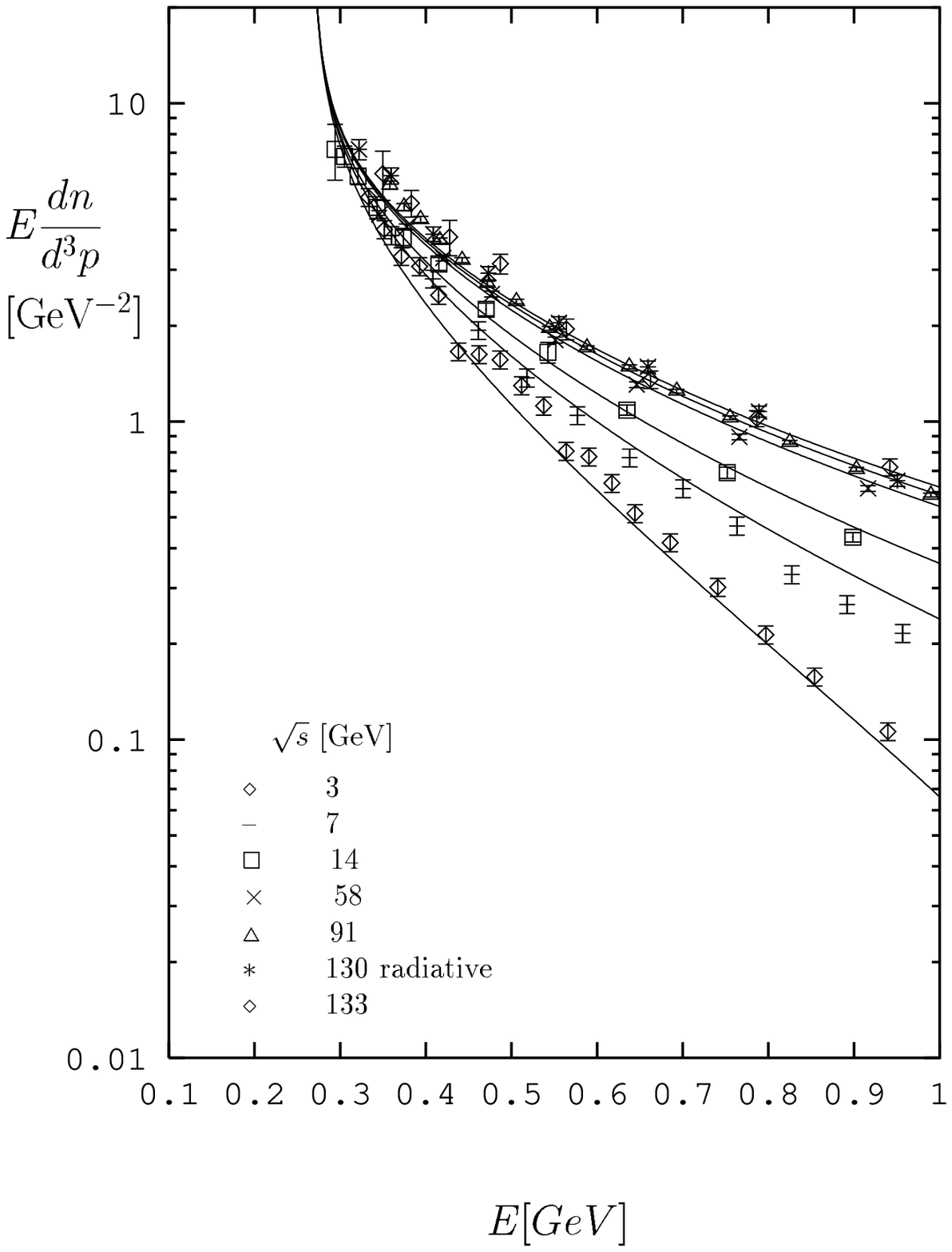,bbllx=4.5cm,bblly=9.5cm,bburx=16.5cm,bbury=26.cm}}
       \end{center}
\caption{Invariant density $E dn/d^3p$ of charged particles
in $e^+e^-$ annihilation 
as a function of the particle energy 
$E=\protect\sqrt{p^2+Q_0^2}$ at $Q_0$ = 270 MeV. 
Data  at various $cms$ energies
are compared to MLLA predictions with the overall normalization adjusted 
(from \protect\cite{klo2}).} 
\end{figure}

\newpage
\begin{thebibliography}{}
%
\bibitem{JETSET}   
 T.\ Sj\"{o}strand, Comp.\ Phys.\ Commun., {\bf
39} (1986) 347; \\  
T.\ Sj\"{o}strand and M.\ Bengtsson, Comp.\ Phys.\ Commun.,
{\bf 43} (1987) 367.

\bibitem{HERWIG}   
G.\ Marchesini and  B.\ R.\ Webber, Nucl.\ Phys.,
{\bf B238} (1984) 1; {\bf 310} (1988) 461;\\
G.\ Marchesini, B.\ R.\ Webber, G.\ Abbiendi, I. G. Knowles, M. H. Seymour,
and L. Stanco,
Comp. Phys. Comm. {\bf 67} (1992) 465.

\bibitem{adkt}   
 Ya.\ I.\ Azimov, Yu.\ L.\ Dokshitzer, V.\ A.\
Khoze and S.\ I.\ Troyan, Z.\ Phys., {\bf C27} (1985) 65 and
{\bf C31} (1986) 213.

\bibitem{dfk1}  
 Yu.\ L.\ Dokshitzer, V.\ S.\ Fadin and V.\ A.\
Khoze, Phys.\ Lett., {\bf 115B} (1982) 242; \\
Z.\ Phys., {\bf C15} (1982) 325.

\bibitem{bcm}   
 A.\ Bassetto, M.\ Ciafaloni and G.\ Marchesini,
Phys.\ Rep., {\bf C100} (1983) 201.

\bibitem{MLLA}
 A.\ H.\ Mueller, Nucl.\ Phys., {\bf B213}
(1983) 85; 
Erratum quoted ibid., {\bf B241} (1984) 141;\\
 Yu.\ L.\ Dokshitzer and S.\ I.\ Troyan, Proc.\
19th Winter School of the LNPI, Vol.\ 1, p.144; Leningrad
preprint LNPI-922 (1984).

\bibitem{dkmt}   
 Yu.\ L.\ Dokshitzer, V.\ A.\ Khoze, A.\ H.\
Mueller and S.\ I.\ Troyan, \lq\lq Basics of Perturbative
QCD", ed.\ J.\ Tran Thanh Van, Editions Fronti\'{e}res,
Gif-sur-Yvette, 1991.

\bibitem{ko} 
for a recent discussion of the phenomenological status of the analytical
perturbative approach, see V. A. Khoze and W. Ochs, `Perturbative QCD
approach to multiparticle production'',
preprint Durham DTP/96/36, MPI-PhT-96/29 (1996).

\bibitem{AO}   
 B.\ I.\ Ermolayev and V.\ S.\ Fadin, JETP.\
Lett., {\bf 33} (1981) 285;\\
 A.\ H. Mueller, Phys.\ Lett., {\bf 104B} (1981)
161.

\bibitem{ahm}
A.\ H. Mueller, Phys. Rev., {\bf D4} (1971) 150.


\bibitem{dfk2}   
 Yu.\ L.\ Dokshitzer, V.\ S.\ Fadin and V.\ A.\
Khoze, Z.\ Phys., {\bf C18} (1983) 37.

\bibitem{ow}   
W.\ Ochs and J.\ Wosiek, Phys.\ Lett.,\ {\bf B304} (1992) 144;
Z.\ Phys.,\ {\bf C68} (1995) 269.

\bibitem{KNO}   
 Z.\ Koba, H.B.\ Nielsen and P.\ Olesen, Nucl. Phys.,
{\bf B40} (1972) 317.
\bibitem{poly}   
 A.M.\ Polyakov,  Sov.\ Phys.\ JETP, {\bf 32} (1971) 296;   
{\bf 33} (1971) 850.

\bibitem{or}
S. J. Orphanidis and V. Rittenberg, Phys. Rev., {\bf D10} (1974) 2892.

\bibitem{bcm1}   
 A.\ Bassetto, M.\ Ciafaloni, G.\ Marchesini,  Nucl.\
Phys., {\bf B163} (1980) 477.

\bibitem{mw}   
 E.\ D.\ Malaza, B.\ R.\ Webber,  Phys.\ Lett.,
{\bf 149B} (1984) 501; Nucl.\ Phys., {\bf B267} (1986) 70.

\bibitem{vietri}   
 W.\ Ochs, Invited talk at the XXIV Int.\ Symp. on
Multiparticle Dynamics, Vietri sul Mare, Italy, Sept.\ 1994,
eds.\ A.\ Giovannini, S.\ Lupia, and R.\ Ugoccioni,
(World Scientific, Singapore), p.243.


\bibitem{bcmm}   
 A.\ Bassetto, M.\ Ciafaloni, G.\ Marchesini and
A.\ H.\ Mueller, Nucl.\ Phys., {\bf B207} (1982) 189.

\bibitem{dktint}   
 Yu.\ L.\ Dokshitzer, V.\ A.\ Khoze and S.\ I.\
Troyan, Int.\ J.\ Mod.\ Phys., {\bf A7} (1992) 1875.

\bibitem{klo1}
V. A. Khoze, S. Lupia and W. Ochs, 
Phys. Lett., {\bf B386} (1996) 451.

\bibitem{lo}   
 S.\ Lupia and W.\ Ochs, Phys.\ Lett., {\bf B365} (1996) 339.

\bibitem{lo2}   
 S.\ Lupia and W.\ Ochs, in preparation.


\bibitem{dkw}   
E.\ A.\ De Wolf, I.\ M.\ Dremin and W.\ K.\ Kittel,
Phys.\ Rep.\ {\bf 270} (1996) 1.

\bibitem{bp}   
A.\ Bia\l as and R.\ Peschanski, Nucl.\ Phys.,\ {\bf B273} (1986) 703;
  \  {\bf B308} (1988) 857. 

\bibitem{dd}   
Yu.\ L.\ Dokshitzer and I.\ Dremin, Nucl.\ Phys.,\ {\bf B402} (1993) 139.

\bibitem{bmp}   
Ph.\ Brax, J.\ L.\ Meunier and R.\ Peschanski, 
Z.\ Phys.,\ {\bf C62} (1994) 649.

\bibitem{ow1}   
W.\ Ochs and J.\ Wosiek, Phys.\ Lett.,\ {\bf B289} (1992) 159.

\bibitem{bm}   
DELPHI Collaboration (preliminary results): 
B.\ Buschbeck, F.\ Mandl\ et al.,
Contributed paper EPS 0553 to 1995 Int.\ Europhysics
 Conference\ on High Energy Physics,
 27 July - 2 August 1995, Brussels, Belgium,
 DELPHI 95-97, PHYS 532 (1995).

\bibitem{corint}
P. Lipa, P. Carruthers, H.C. Eggers and B. Buschbeck, Phys. Lett.,
{\bf B285} (1992) 300.
\bibitem{dmo}   
Yu.\ L.\ Dokshitzer, G.\ Marchesini and
G.\ Oriani, Nucl.\ Phys., {\bf B387} (1992) 675.  

\bibitem{ow3}   
W.\ Ochs and J.\ Wosiek, Z.\ Phys., {\bf C72} (1996) 263.

\bibitem{vakcar}   
 V.\ A.\ Khoze, in  in \lq\lq $Z^0$ Physics",
Carg\`{e}se 1990, Eds.\ M.\ L\'{e}vy, J-L.\ Basdevant, M.\ Jacob,
D.\ Speiser, J.\ Weyers and R.\ Gastmans, NATO ASI, Series B:
Physics Vol.\ 261, p. 419.


\bibitem{klo2}
V.\ A.\ Khoze, S.\ Lupia and W.\ Ochs, ``QCD coherence and the soft limit of
the energy spectrum", preprint DTP/96/90, MPI-PhT/96-92, hep-ph/9610204,
October 1996.
%
\bibitem{klo3}
V.\ A.\ Khoze, S.\ Lupia and W.\ Ochs, 
``Soft particle production and QCD coherence",
preprint DTP/96/94, MPI-PhT/96-94, hep-ph/9610348, October 1996.

\end{thebibliography}
\end{document}